\RequirePackage{fix-cm}
\documentclass[smallextended]{svjour3}       % onecolumn (second format)
\smartqed  % flush right qed marks, e.g. at end of proof
\usepackage{graphicx}
\usepackage{enumerate}
\begin{document}

\title{Time-domain digital-to-analog converter for spiking neural network hardware%\thanks{Grants or other notes
}
\subtitle{}


\author{Seiji Uenohara         \and
        Kazuyuki Aihara %etc.
}


\institute{S. Uenohara \at
              Institute of Industrial Science, The University of Tokyo 4-6-1 Komaba, Meguro-ku, Tokyo 153-8505, Japan \\
             % Tel.: +123-45-678910\\
             % Fax: +123-45-678910\\
              \email{s.uenohara@gmail.com}            \\
          % \and
           K. Aihara \at
           Institute of Industrial Science, The University of Tokyo 4-6-1 Komaba, Meguro-ku, Tokyo 153-8505, Japan \\
           International Research Center for Neurointelligence, University of Tokyo 7-3-1 Hongo, Bnkyo-ku, Tokyo 113-8654, Japan \\
           \email{aihara@sat.t.u-tokyo.ac.jp}
}

\date{Received: date / Accepted: date}

\maketitle

\begin{abstract}
We propose a new digital-to-analog converter~(DAC) for 
realizing a synapse circuit of mixed-signal spiking neural networks.
We named this circuit ``time-domain DAC~(TDAC)''.
This produces weights for converting a digital input code into voltage using one current waveform.
Therefore, a TDAC is more compact than a conventional DAC that comprises many current sources and resistors.
Moreover, a TDAC with leak resistance reproduces biological plausible synaptic responses 
that are expressed as alpha functions or dual exponential equations.
We will show  numerical analysis results of a TDAC and 
circuit simulation results of a circuit designed by the TSMC 40~nm CMOS process.

\end{abstract}

\section{Introduction}
\label{intro}
Application-specific integrated circuits~(ASICs) for neuromorphic hardware 
have been studied intensively with the aim of achieving highly efficient 
computation~\cite{schemmel2008wafer,tanaka2009cmos,truenorth2015,rolls2015,indiveri2015neuromorphic,stanford2018}.
Although most neuromorphic hardware is designed as digital circuits~\cite{truenorth2015,rolls2015},
some studies have sought higher efficiency by using analog circuits~\cite{indiveri2015neuromorphic} due to
the advantage that highly efficient multiply-accumulate~(MAC) operations can be achieved.

Neuromorphic computation requires many MAC operations for the input signals and synaptic weights,
and therefore implementation of highly efficient MAC operations is important for realizing the highly efficient neuromorphic hardware. 
To improve the energy efficiency of MAC operations, attempts have been made to 
(i) binarize the synaptic weights, 
(ii) realize weighted summation using current~(wired-sum),
(iii) realize weighted summation using capacitors, 
and (iv) develop a method that combines (i)--(ii).

Bankman et al.~\cite{stanford2018} realized a highly energy-efficient binary convolution neural network chip
to act as a recognition processor.
In this circuit, the synaptic weights express binary information and are stored in digital memory. 
The weighted summation  is then calculated by converting the binary information into analog voltages. 
The capacitors on this chip for weighted summation work also as a digital-to-analog converter~(DAC), 
and this circuit is categorized as a mixed-signal circuit. 
A mixed--signal circuit that has multi-bit synaptic weights requires 
a large footprint area and a large amount of energy for the DAC, 
but the chip developed by Bankman et~al. has low power consumption 
because the synaptic weights are restricted to be binary 
and the weighted summation is realized by capacitance coupling.

Weighted summation by current is often used in analog neuromorphic hardware~\cite{tanaka2009cmos,indiveri2015neuromorphic}.
In such circuits, the current is produced by synapse circuits. Current-based summation is 
implemented very simply by connecting each of the metal lines of the synaptic output.
In analog neuromorphic hardware, synaptic weights are often expressed 
by analog voltages that are maintained by capacitors. A circuit with this architecture is highly efficient, 
but it is difficult to reuse the capacitors that hold the synaptic weights 
because the former cannot hold the weights for long due to charge leakage. 
To solve this problem, there have been many studies of 
synapse circuits that use analog memory~\cite{seo2011,wang2015,adhikari2012,burr2017,kuzum2013}.
However, there are many problems to be solved for establishing analog memory as reliable technology.

To realize highly efficient neuromorphic hardware with reuseable synaptic weights
fabricated using conventional complementary metal--oxide--semiconductor~(CMOS) technology,
a circuit architecture synaptic weights are kept by digital memories and MAC operations are achieved by an analog circuit, 
i.e. a mixed-signal architecture is suitable. 
Realizing a mixed-signal circuit that has multi-bit synaptic weights is important 
for achieving on-chip learning, but it is difficult to realize 
high-integration and highly efficient neuromorphic hardware 
because conventional multi-bit DACs comprise many current-source circuits 
or resistor arrays, thereby necessitating a large footprint 
and high power consumption~\cite{miki,post}. It is especially difficult to 
implement a highly energy-efficient asynchronous spiking neural network~(SNN) 
chip that has multi-bit synaptic weights because using one 
DAC per time division is difficult in an asynchronous system.

To realize high-integration and highly energy-efficient SNN hardware with on-chip learning, 
we propose a new DAC circuit that weights each bit of digital memory 
by using a current (or voltage) waveform. This makes our DAC more compact than a conventional one, 
and we refer to this DAC as a ``time-domain'' DAC~(TDAC).
Herein, we present the results of numerical analysis of TDAC and circuit simulation.
  
This paper is organized as follows: In Section~2, we explain the principle of TDAC, and in Section~3, 
numerical analysis. Circuit design of TDAC and circuit simulation results are shown in Section~4.
Finally, we conclude this paper in Section~6.

\section{Circuit principle of time-domain digital-to-analog converter}

\begin{figure}
\includegraphics[width=1.0\textwidth]{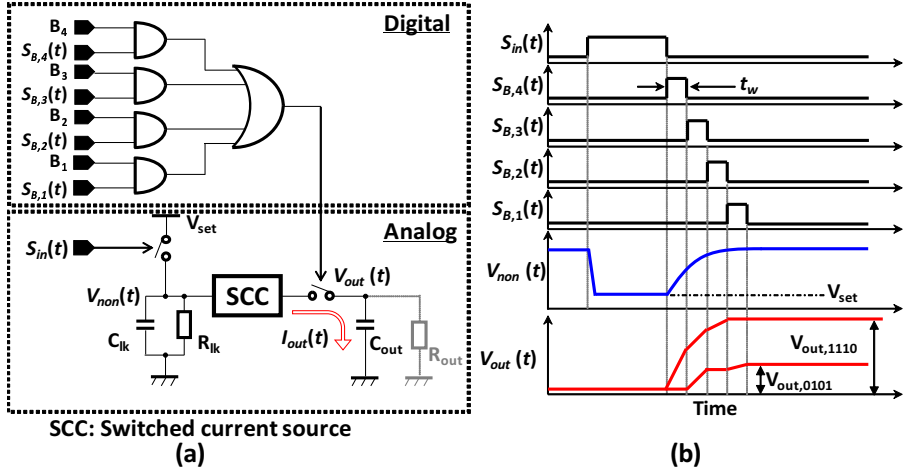}
\caption{Principle of time-domain digital-to-analog converter (TDAC) in circuit form: 
(a) circuit; (b) voltage waveform of each node of circuit in (a).}
\label{fig:prc}  
\end{figure}

Figure~\ref{fig:prc}(a) shows the TDAC principle in circuit form, and we explain the operation 
of a four-bit TDAC as an example. The TDAC consists of an analog block and a digital block 
that comprise AND gates, an OR gate, switches, a switched current source~(SCC), 
resistors, and capacitors . The digital memory values are $B_1$--$B_4$, and in this example, 
$B_1$ and $B_4$ are the least significant bit~(LSB) and the most significant bit~(MSB), respectively. 
$S_{in}(t)$ is the trigger signal for activating the DAC and corresponds to spike pulses 
when the TDAC is used as the output stage of a synapse circuit. 
Signals $S_{B,1}(t)$--$S_{B,4}(t)$ are non-overlapping digital signals with pulse width $t_w$ for the DAC, 
and these values are either zero or unity. The SCC outputs a current $I_{out}(t)$ ($\propto V_{non}(t)$), 
and the leak resistance $R_{out}$ is an option for realizing the waveform of synaptic potential.

The DAC process using the TDAC without the leak resistance $R_{out}$ is as follows (see Fig.~\ref{fig:prc}(b)):
%-=-=-=-=-=-=-=-=-=-=-=-=-=-=-=-=-=-=-=-=-=-=-=-=-=-=-=-=-=-=-=-=-=-=-=-=-=-=-=-=-=-=
\begin{enumerate}[1)]
\item When $S_{in}(t)$ is high, $V_{non}(t)$ is set to $\rm{V_{set}}$.
\item When $S_{in}(t)$ is turns off, $S_{B,4}(t)$ is generated at the trailing edge of $S_{in}(t)$. 
At the same time, $V_{non}(t)$ increases exponentially with time constant $C_{lk}R_{lk}$.
If $B_{4}$ is high, then capacitor $C_{out}$ is charged during $t_w$ by a current proportional to $V_{non}(t)$.
\item $S_{B,3}(t)$ is generated at the trailing edge of $S_{B,4}$. If $B_{3}$ is high, capacitor $C_{out}$ is charged during $t_w$ by a current that is proportional to $V_{non}(t)$.
\item Operation with 3) is repeated until $S_{B,1}(t)$ is generated.
\end{enumerate}
%-=-=-=-=-=-=-=-=-=-=-=-=-=-=-=-=-=-=-=-=-=-=-=-=-=-=-=-=-=-=-=-=-=-=-=-=-=-=-=-=-=-=
For example, $V_{out,1110}$ represents the voltage-converted digital input code $(1110)_2$ (see Fig. \ \ref{fig:prc}(b))
that is obtained by charging, charging, charging, and not charging.

Conventional DACs use many resistors or current sources to weight each bit of digital memory. Implementing these circuit components causes that the footprint area must be squared for every unit increase in the length of the memory.
By contrast, a TDAC uses a current waveform to weight each memory bit. In the TDAC, the numbers of AND gates, OR gates, and signal generators of $S_{B,k}$ needed to sample the current waveform increases as the length of the memory increases.
Therefore, the number of transistors increases linearly, but the disadvantage is that the time taken by the DAC increases. A TDAC is suitable for hardware that must be compact and operate at low speed, such as SNN hardware. In this example, the output current is positive but the TDAC can output negative current.
%-------------------------------------------
%-------------------------------------------

\section{Numerical analysis of time-domain digital-to-analog converter}

\subsection{Time-domain digital-to-analog conversion without leak resistance}

We define the time of the trailing edge of $S_{in}(t)$ to be zero, and we express the voltage $V_{out}(t)$ as 
%====================================================================================%
\begin{equation}
V_{out}(t) = \sum^{q-1}_{k=0} \int_{kt_w}^{(k+1)t_{w}} B_{q-k}\frac{f_{scc}(V_{non}(t))}{C_{out}} \mathrm{d}t, \label{eq:vdac}
\end{equation}
%====================================================================================%
where $q$ is the memory length, $t$ is continuous time ($t \geq 0$), and $f_{scc}(\cdot)$ characterizes the SCC. In this section, we analysis the TDAC when $V_{non}(t)$ is described by $V_{set}\exp(-\frac{t}{C_{lk}R_{lk}})$ and $f_{scc}(V_{non}(t))=V_{non}(t)$. Under these conditions, Eq.~(\ref{eq:vdac}) is expressed as 
%====================================================================================%
\begin{equation}
V_{out}(t) = \frac{V_{set}}{C_{out}}\sum^{q-1}_{k=0} \int_{kt_w}^{(k+1)t_{w}} B_{q-k}\exp(-\frac{t}{C_{lk}R_{lk}}) \mathrm{d}t. \label{eq:vdac2}
\end{equation}
%====================================================================================%
We integrate Eq.~(\ref{eq:vdac2}) to obtain 
%====================================================================================%
\begin{equation}
V_{out}(t) = \frac{V_{set}}{C_{out}}\sum^{q-1}_{k=0} -B_{q-k}C_{lk}R_{lk}(\exp(-\frac{(k+1)t_w}{C_{lk}R_{lk}}) - \exp(-\frac{kt_w}{C_{lk}R_{lk}}) ). \label{eq:vdac3}
\end{equation}
%====================================================================================%

To ensure that the DAC characteristics remain linear, the weight of the upper bit of the adjacent must be twice that of the lower bit, namely 
%====================================================================================%
\begin{equation}
\frac{(\exp(-\frac{(k+1)t_w}{C_{lk}R_{lk}}) - \exp(-\frac{kt_w}{C_{lk}R_{lk}}) )}{(\exp(-\frac{(k+2)t_w}{C_{lk}R_{lk}}) - \exp(-\frac{(k+1)t_w}{C_{lk}R_{lk}}) )} = 2.
\end{equation}
%====================================================================================%
We solve Eq. (4) to obtain 
%====================================================================================%
\begin{equation}
\frac{t_w}{C_{lk}R_{lk}} = \ln 2.
\end{equation}
%====================================================================================%

\begin{figure}
\includegraphics[width=0.75\textwidth]{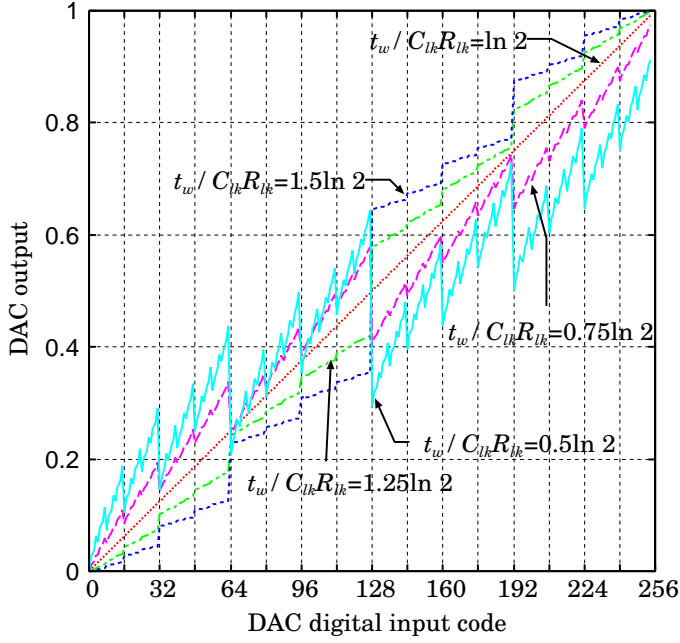}
\caption{Input-output characteristics of TDAC obtained 
from numerical simulation for varying $\frac{t_w}{C_{lk}R_{lk}}$.}
\label{fig:io_sim}       
\end{figure}

Figure~\ref{fig:io_sim} shows the input-output characteristics of 
the TDAC obtained from numerical simulation for varying $\frac{t_w}{C_{lk}R_{lk}}$. 
As shown therein, the characteristics are linear when $\frac{t_w}{C_{lk}R_{lk}} = \ln 2$ 
but nonlinear otherwise.
 In particular, monotonicity is also lost when $\frac{t_w}{ C_{lk}R_{lk} } < \ln 2$.
%-------------------------------------------
%-------------------------------------------

\subsection{Time-domain digital-to-analog conversion with leak resistance}

When a TDAC is used as the output stage of a synaptic circuit, we can view $C_{out}$ as being the membrane capacitance of an analog neuron circuit, where the leak resistance $R_{out}$ is connected in parallel with $C_{out}$. In this situation, the temporal variation of $V_{out}(t)$ with $R_{out}$ is expressed by 
%====================================================================================%
\begin{equation}
\frac{dV_{out}(t)}{dt} = -\frac{V_{out}(t)}{C_{out}R_{out}}+V_{set}\exp(-\frac{t}{C_{lk}R_{lk}})\sum^{q-1}_{k=0} S_{B,q-k}(t)B_{q-k}. \label{eq:lkv1}
\end{equation}
%====================================================================================%
We will show that Eq.~(\ref{eq:lkv1}) fits well with biological data on synaptic potentials when all the memory bits are unity 
and $qt_w$ is sufficient large, in which case $\sum^{q-1}_{k=0} S_{B,q-k}(t)B_{q-k}$ is nearly unity regardless of $t$. We assume that $V_{out}(0)$ is zero, and in this case Eq.~(\ref{eq:lkv1}) is expressed as 
%====================================================================================%
\begin{equation}
\frac{dV_{out}(t)}{dt} = -\frac{V_{out}(t)}{C_{out}R_{out}}+V_{set}\exp(-\frac{t}{C_{lk}R_{lk}}). \label{eq:lkv2}
\end{equation}
%====================================================================================%
We solve  Eq.~(\ref{eq:lkv2}) by using the method of variation of constants, setting $C_{out}R_{out}=\tau_1$ and $C_{lk}R_{lk}=\tau_2$. The solution of Eq.~(\ref{eq:lkv2}) is given by
%====================================================================================%
\begin{equation}
V_{out}(t) = g(t)\exp(\frac{-t}{\tau_1}), \label{eq:lkv_} 
\end{equation}
%====================================================================================%
where $g(\cdot)$ is a function of time. By using Eqs.~(\ref{eq:lkv1}) and (\ref{eq:lkv2}), $\frac{dV_{out}(t)}{dt}$ is expressed as 
%====================================================================================%
\begin{eqnarray}
\frac{dV_{out}(t)}{dt} = g'(t)\exp(-\frac{t}{\tau_1}) - \frac{1}{\tau_1}g(t)\exp(-\frac{t}{\tau_1}), \nonumber \\
g'(t) = V_{set}\exp({-\frac{\tau_1-\tau_2}{\tau_1\tau_2}t}). \label{eq:g'}
\end{eqnarray}
%====================================================================================%
If $\tau_1=\tau_2$, then the integral of Eq.~(\ref{eq:g'}) with respect to $t$ is $tV_{set}$ because $g'(t)$ is a constant. In this case, $V_{out}(t)$ is given by 
%====================================================================================%
\begin{equation}
V_{out}(t) =  tV_{set}\exp(\frac{-t}{\tau_1}), \label{eq:alpha} 
\end{equation}
%====================================================================================%
which is an alpha function. 

If $\tau_1\neq\tau_2$, then the integral of Eq.~(\ref{eq:g'}) with respect to $t$ is 
%====================================================================================%
\begin{equation}
g(t) = -\frac{\tau_1\tau_2}{\tau_1-\tau_2}V_{set}\exp(-\frac{\tau_1-\tau_2}{\tau_1\tau_2}t) + g_c,
\end{equation}
%====================================================================================%
where $g_c$ is a constant of integration. By using $V_{out}(0)=0$, we obtain $g_c = \frac{\tau_1\tau_2}{\tau_1-\tau_2}V_{set}$. Substituting $g_c$ for $g(t)$, we obtain 
%====================================================================================%
\begin{equation}
g(t) = -\frac{\tau_1\tau_2}{\tau_1-\tau_2}V_{set} (1-\exp(-\frac{\tau_1-\tau_2}{\tau_1\tau_2}t)). \label{eq:gt}
\end{equation}
%====================================================================================%
Substituting Eq.~(\ref{eq:gt}) for Eq.~(\ref{eq:lkv_}), we obtain 
%====================================================================================%
\begin{equation}
V_{out}(t) = \frac{\tau_1\tau_2}{\tau_1-\tau_2}V_{set}(\exp(-\frac{t}{\tau_1}) - \exp(-\frac{t}{\tau_2})), \label{eq:dual}
\end{equation}
%====================================================================================%
which is a dual exponential function. It is known that alpha and dual exponential functions fit well to biological data on synaptic potentials~\cite{otis1993,rall1967,wilson1992}.

\begin{figure}
\includegraphics[width=1.0\textwidth]{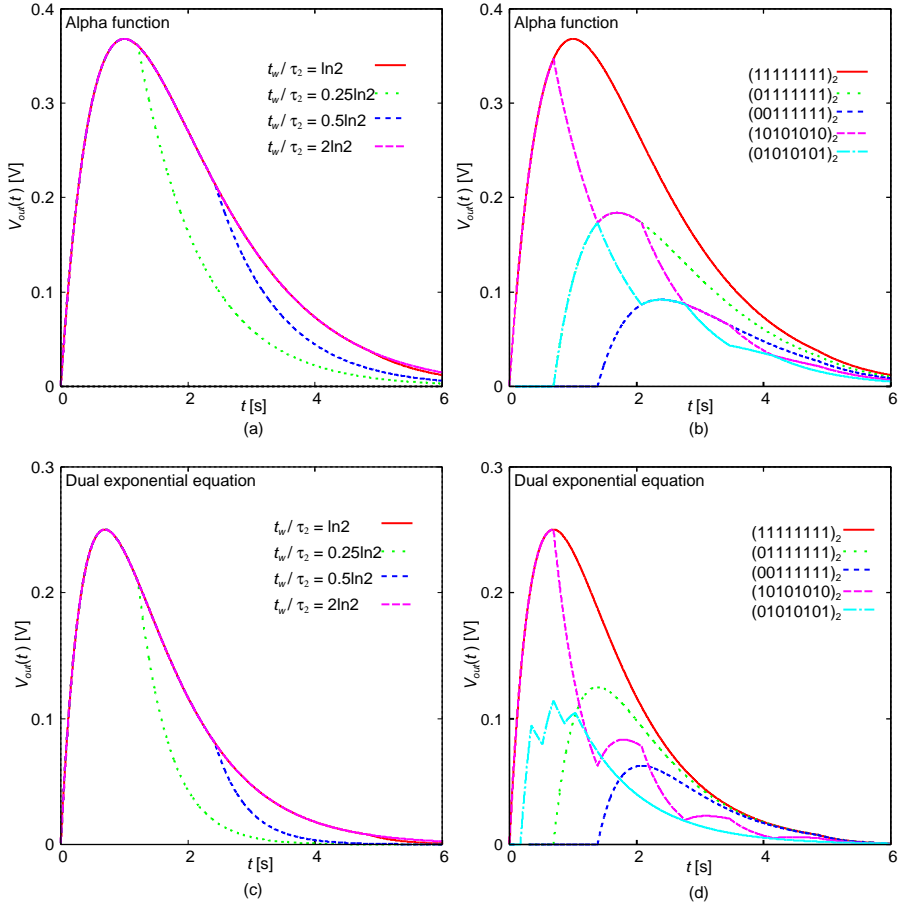}
\caption{Synaptic potential waveforms obtained from numerical simulation: (a) $\tau_1=1$ and $\tau_2=1$ when $\frac{t_w}{\tau_2}$ is varied; (b) $\tau_1=1$ and $\tau_2=1$ when the DAC digital input code is varied; (c) $\tau_1=1$ and $\tau_2=0.5$ when $\frac{t_w}{\tau_2}$ is varied; (d) $\tau_1=1$ and $\tau_2=0.5$ when the DAC digital input code is varied.}
\label{fig:pspsim}       
\end{figure}

Figure~\ref{fig:pspsim} shows synaptic potential waveforms obtained from numerical simulation. Panels (a) and (b) show the waveforms with the alpha function ($\tau_1=\tau_2$), and (c) and (d) show those with the dual exponential function ($\tau_1\neq\tau_2$). To obtain panels (a) and (c) and panels (b) and (d), we changed $\frac{t_w}{\tau_2}$ and the digital input code, respectively. In panels (a) and (c), the waveforms are so similar that they cannot be distinguished visually, and the peak of the potential does not change when $\frac{t_w}{\tau_2}$ is varied. By contrast, as shown in panels (b) and (d), when the input code is $(10101010)_2$ or $(01010101)_2$, the peak changes. In the TDAC, $C_{out}$ is not charged during $S_{B,k}$ when memory bit $k$ is zero, and this is caused by alternating charging and leaking. Moreover, the upper bit is converted into analog current faster than the lower one. 
However, we can ignore the influence of these when $qt_w$ is sufficiently smaller than the time constant of the membrane potential of a neuron circuit.

%-------------------------------------------
%-------------------------------------------

\section{Proposed circuit}

\begin{figure}
\includegraphics[width=1.0\textwidth]{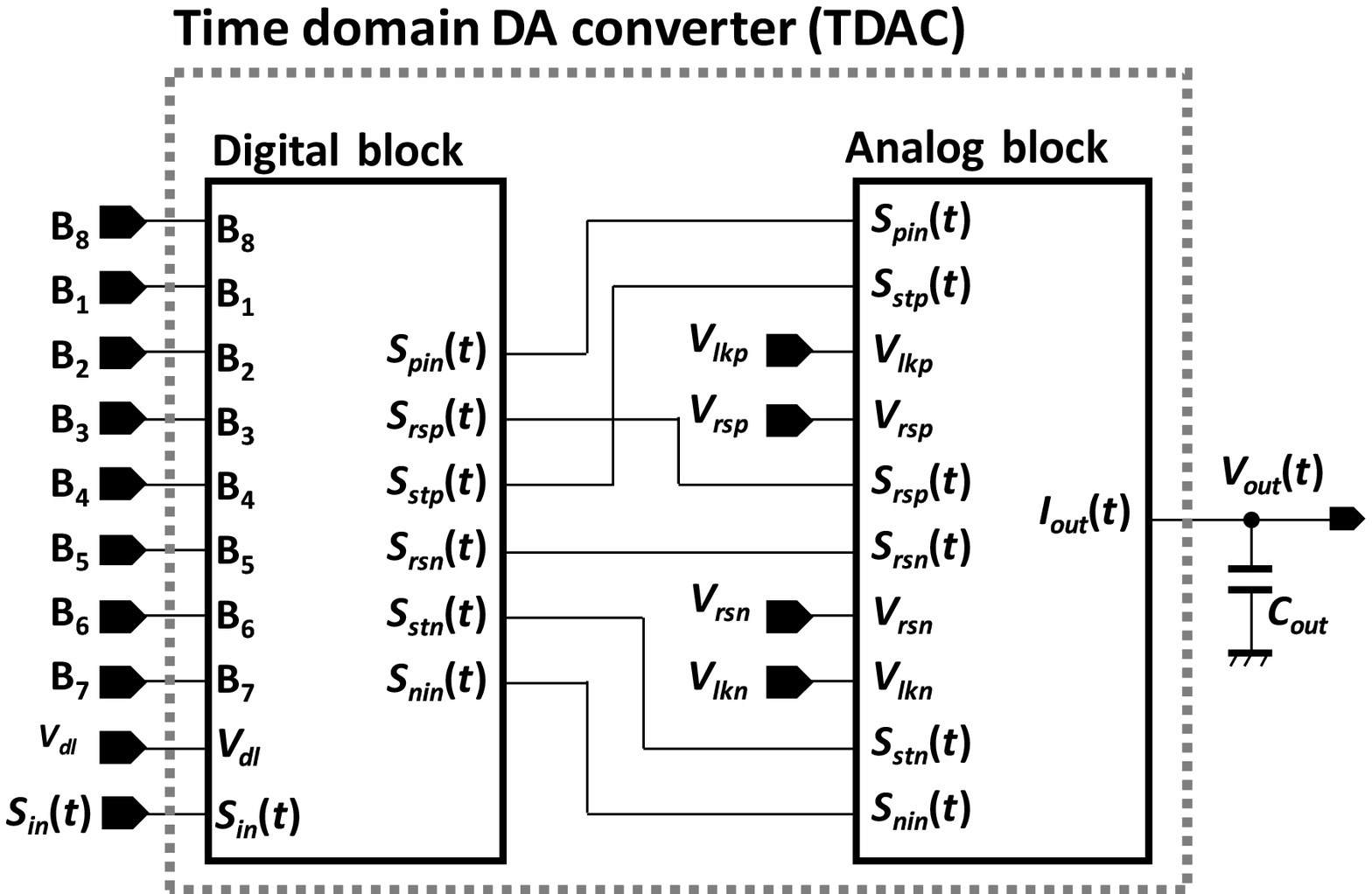}
\caption{Block diagram of proposed eight-bit TDAC.}
\label{fig:tdac}
\end{figure}

\begin{figure}
\includegraphics[width=1.0\textwidth]{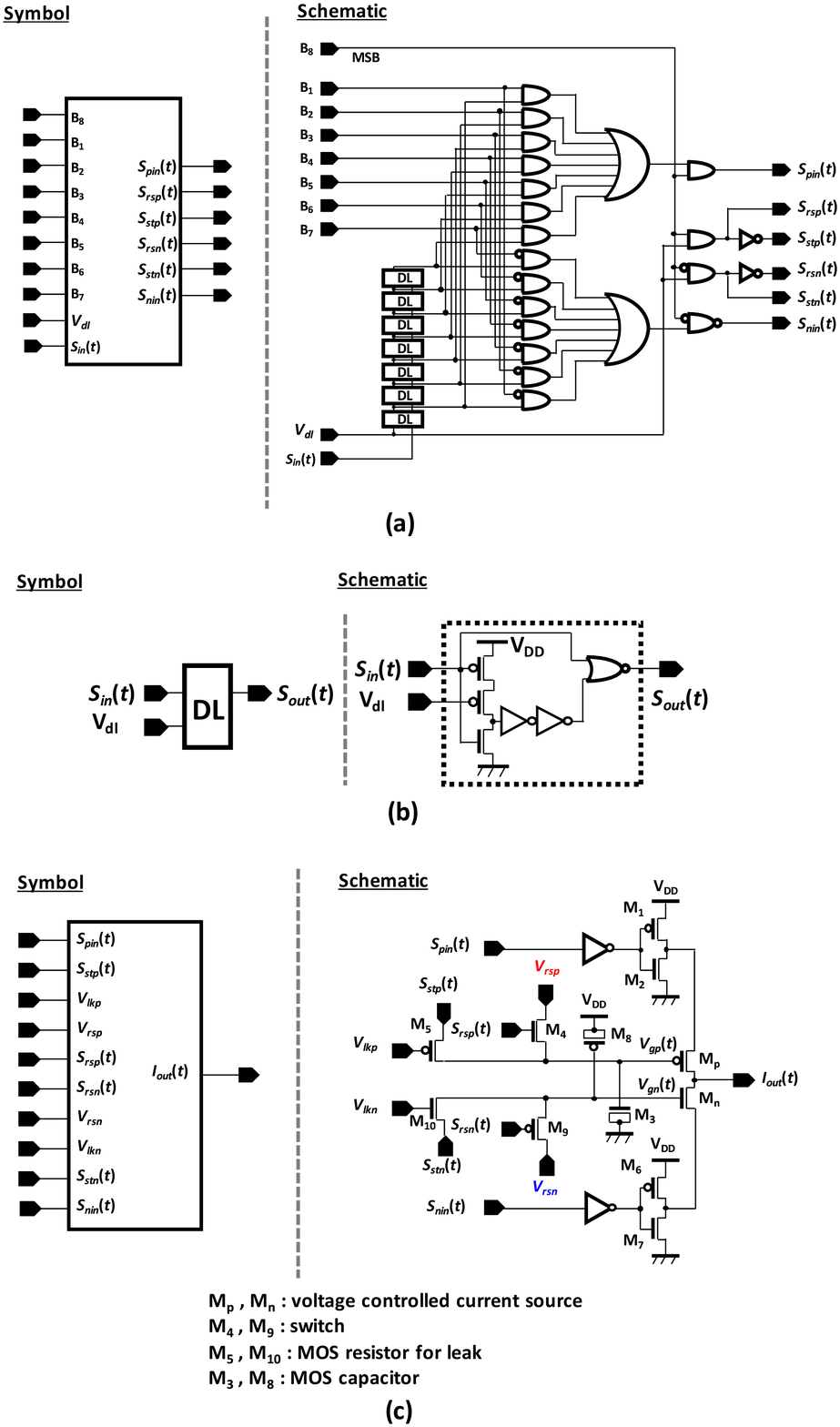}
\caption{Details of (a) digital block, (b) delay circuit, and (c) analog block of proposed circuit.}
\label{fig:detail}
\end{figure}

A block diagram of our proposed circuit is shown in Fig.~\ref{fig:tdac}, where an eight-bit (sign+7) DAC can output negative and positive current. The circuit consists of a digital block and an analog block, the details of which are shown in Fig.~\ref{fig:detail}.
%-------------------------------------------
%-------------------------------------------

The digital block consists of logic gates and delay circuits, DLs. Each DL consists of 11~transistors that output $S_{B,k}$ with pulse width $t_w$. The pulse width is the adjusted bias voltage $V_{dl}$.
If the bit-length of the TDAC increases, then an additional DL, two NAND gates, a NOT gate, two NMOS transistors, and two PMOS transistors are required, thereby increasing the total number of transistors by 25. The output signal of the digital block is the input for the analog block.

The analog block consists of voltage-controlled current sources, switches, MOS resistors, and MOS capacitors. Negative and positive current outputs are realized by transistors $\rm{M_{n}}$ and $\rm{M_{p}}$, respectively. The number of transistors in the analog block does not increase when the bit-length of the TDAC increases. We explain the operation of positive and negative current output in Sections~4.2 and 4.3, respectively. 
%-------------------------------------------
%-------------------------------------------

\subsection{Positive current output}

The proposed circuit outputs positive current when $B_{8}$ is unity. As an example, Fig.~\ref{fig:tm}(a) shows the timing diagram of the input signal and node voltages when the input digital bit code is $(11101000)_2$. The process for outputting positive current is as follows.
%-=-=-=-=-=-=-=-=-=-=-=-=-=-=-=-=-=-=-=-=-=-=-=-=-=-=-=-=-=-=-=-=-=-=-=-=-=-=-=-=-=-=
\begin{enumerate}[1)]
\item When $S_{in}(t)$ turns high, $S_{rsp}(t)$ and $S_{stp}(t)$ turn high and low, respectively. At the same time, $V_{gp}(t)$ is set to $V_{rsp}$, where $V_{rsp}$ is a small voltage.
\item When $S_{in}(t)$ turns off, $S_{B,7}(t)$ is generated at the trailing edge of $S_{in}(t)$. At the same time, $V_{gp}(t)$ increases exponentially.
\item $S_{B,6}(t)$ is generated at the trailing edge of $S_{B,7}$. If $B_{6}$ is high, then capacitor $C_{out}$ is charged during $t_w$ by $\rm{M_{p}}$.
\item Operation~3 is repeated until $S_{B,1}(t)$ is generated.
\end{enumerate}
%-=-=-=-=-=-=-=-=-=-=-=-=-=-=-=-=-=-=-=-=-=-=-=-=-=-=-=-=-=-=-=-=-=-=-=-=-=-=-=-=-=-=
We prevent wasteful power consumption by setting $S_{stp}(t)$ to low while $S_{in}(t)$ is high. This is because wasteful current flows from $\rm{M_5}$ to $\rm{M_4}$ if $S_{stp}(t)$ is high while $S_{in}(t)$ is high.

\subsection{Negative current output}

\begin{figure*}
\includegraphics[width=1.0\textwidth]{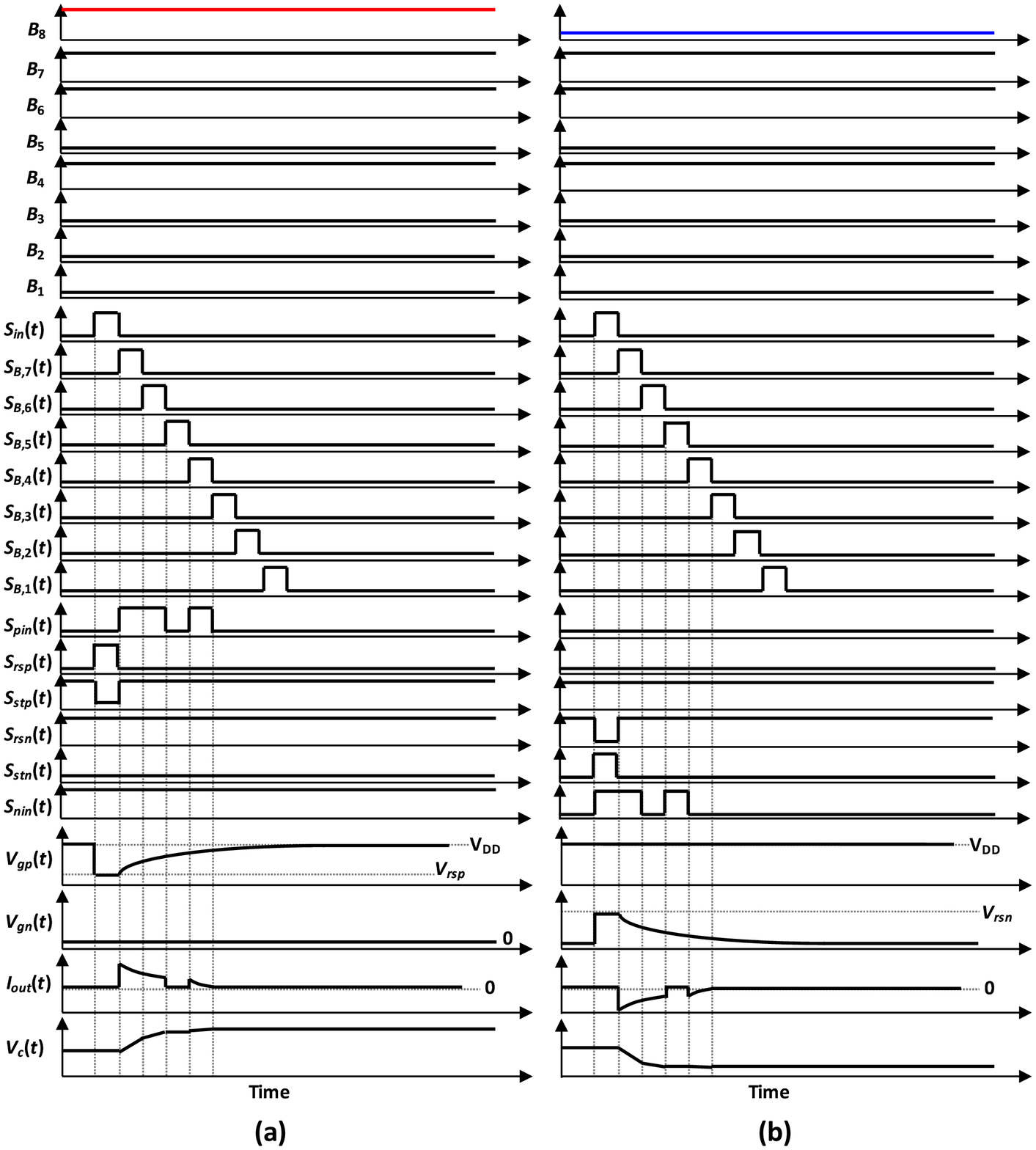}
\caption{Timing diagrams of input signal and node voltages for (a) positive and (b) negative current output operations.}
\label{fig:tm}
\end{figure*}

The proposed circuit outputs negative current when $B_{8}$ is zero. Figure~\ref{fig:tm}(b) shows the timing diagrams of the input signal and node voltages when the input digital bit code is $(01101000)_2$. The process for outputting positive current is as follows.
%-=-=-=-=-=-=-=-=-=-=-=-=-=-=-=-=-=-=-=-=-=-=-=-=-=-=-=-=-=-=-=-=-=-=-=-=-=-=-=-=-=-=
\begin{enumerate}[1)]
\item When $S_{in}(t)$ turns high, $S_{rsn}(t)$ and $S_{stn}(t)$ turn low and high, respectively. At the same time, $V_{gn}(t)$ is set to $V_{rsn}$, where $V_{rsn}$ is a high voltage.
\item When $S_{in}(t)$ turns off, $S_{B,7}(t)$ is generated at the trailing edge of $S_{in}(t)$. At the same time, $V_{gn}(t)$ decreases exponentially.
\item $S_{B,6}(t)$ is generated at the trailing edge of $S_{B,7}$. If $B_{6}$ is high, then capacitor $C_{out}$ is charged during $t_w$ by $\rm{M_{n}}$.
\item Operation~3 is repeated until $S_{B,1}(t)$ is generated.
\end{enumerate}
%-=-=-=-=-=-=-=-=-=-=-=-=-=-=-=-=-=-=-=-=-=-=-=-=-=-=-=-=-=-=-=-=-=-=-=-=-=-=-=-=-=-=
We prevent wasteful power consumption by setting $S_{stn}(t)$ to high while $S_{in}(t)$ is high. This is because wasteful current flows from $\rm{M_9}$ to $\rm{M_{10}}$ if $S_{stn}(t)$ is low while $S_{in}(t)$ is high.
%-------------------------------------------
%-------------------------------------------

\section{Circuit simulation}

\subsection{Time-domain digital-to-analog conversion without leak resistance}

We designed an eight-bit TDAC as shown in Fig.~\ref{fig:tdac} with the TSMC 40~nm CMOS process~(1~poly, 8~metal), and we evaluated the circuit by means of the Spectre simulation. We set the bias voltages and the capacitance as $\rm{V_{DD}}=700$~mV, $V_{dl}=180$~mV, $V_{lkp}=140$~mV, $V_{lkn}=420$~mV, and $C_{out}=0.5$~pF, and the output voltage $V_{out}(t)$ was reset to 350~mV on every input.

\begin{figure}
  \includegraphics[width=0.8\textwidth]{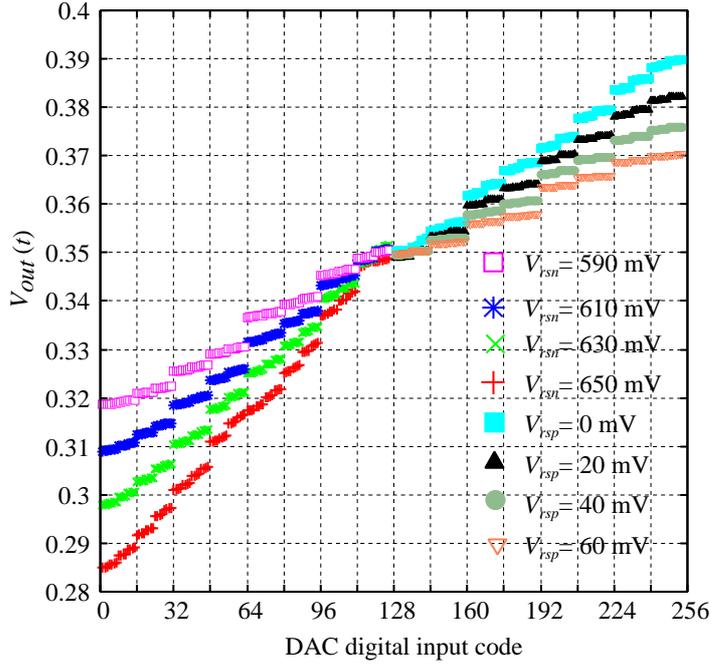}
\caption{Input--output characteristics of TDAC.}
\label{fig:io}       % Give a unique label
\end{figure}

Figure~\ref{fig:io} shows the input--output characteristics of the TDAC when $V_{rsn}$ and $V_{rsp}$ are varied separately. In the designed eight-bit TDAC, the output current is negative when the digital input code is between zero and 128, and it is positive when the digital input code is between 129 and 255. The slopes of the characteristics in the two regions can be adjusted separately by varying $V_{rsn}$ and $V_{rsp}$ as shown in Fig.~\ref{fig:io}. The energy per one digital-to-analog conversion is 27~fJ when the digital input code is $(11111111)_2$ and $V_{rsp}=0$~mV.

%-------------------------------------------
%-------------------------------------------
 
\subsection{Time-domain digital-to-analog conversion with leak resistance}

The circuit simulation for synaptic-potential generation was conducted by adding a MOS resistance between the output node and the ground. We set the bias voltages and the capacitance as $\rm{V_{DD}}=700$~mV, $V_{dl}=340$~mV, $V_{lkp}=300$~mV, $V_{lkn}=300$~mV, $V_{rstn}=570$~mV, $V_{rstp}=70$~mV, and $C_{out}=0.5$~pF.

\begin{figure}
  \includegraphics[width=0.8\textwidth]{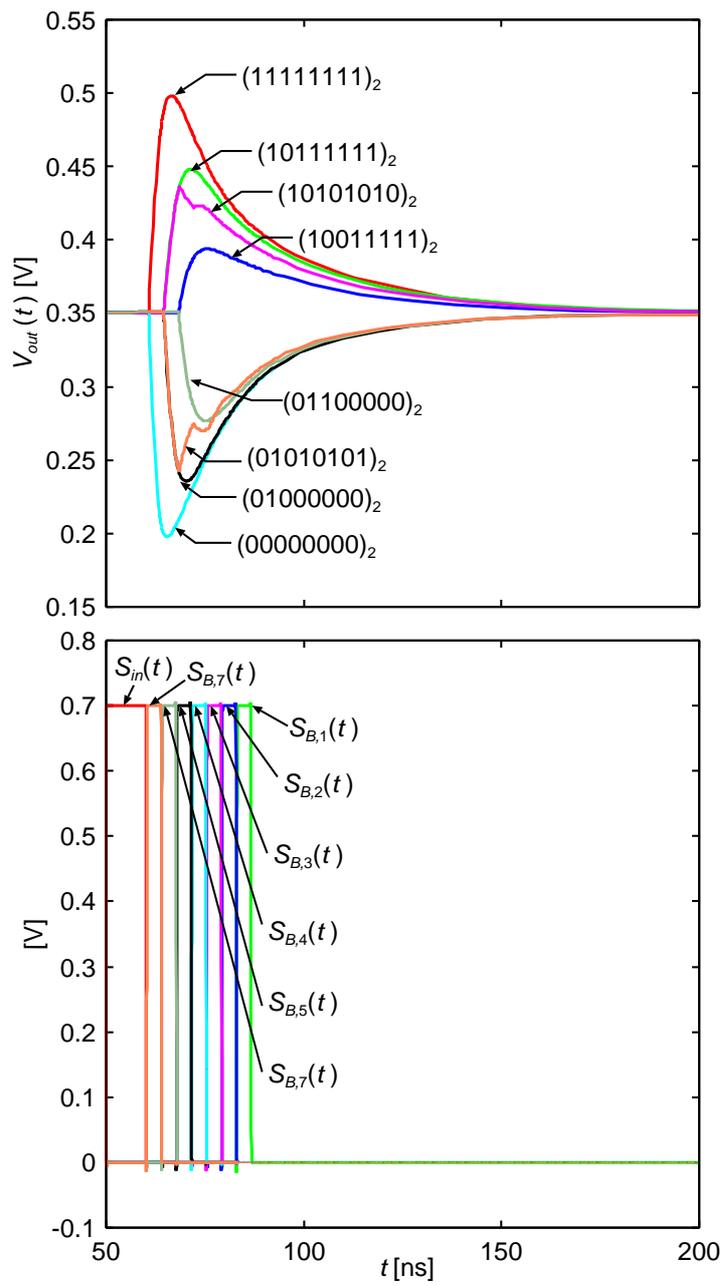}
\caption{Synaptic-potential waveforms obtained from circuit simulation for various DAC digital input codes.}
\label{fig:pspcir}       % Give a unique label
\end{figure}

Figure~\ref{fig:pspcir} shows the synaptic-potential waveforms for various digital input codes $S_{in}(t)$ and DL outputs. As shown in Fig.~\ref{fig:pspsim}, which was obtained by numerical simulation, the waveforms are smooth when the lined bit is unity or zero, but in other cases the waveform has multiple peaks. We obtained waveforms that are similar to those from the numerical simulation.
%-------------------------------------------
%-------------------------------------------

\section{Conclusion}

We proposed a new DAC, or a TDAC, in which the weight of each bit 
that codes for the DAC is realized by a current waveform sampled using non-overlapping 
digital signals. The number of transistors needed to implement a TDAC increases 
linearly with the bit-length, but the transistors can be made small because they work as a digital circuit. 
A TDAC is therefore more compact than a conventional DAC in which the number of 
transistors increases exponentially with the bit-length. Moreover, a TDAC with leak 
resistance realizes biologically plausible synaptic responses without the need 
for other circuit components. TDACs are therefore suitable for implementing mixed-signal 
SNN hardware that requires high integration.

We showed the condition under which a TDAC remains linear, namely that the ratio of the pulse 
width for sampling the current waveform to the time constant should be $\ln 2$. 
To realize a TDAC that has good linearity and is robust against fabrication mismatches, 
the pulse width (resp.\ time constant) should be set according to the time constant (resp.\ pulse width), 
and we intend to develop such a circuit in our future work.

\begin{acknowledgements}
This work is supported by VLSI Design and Education
Center(VDEC), the University of Tokyo in collaboration
with Cadence Design Systems, Inc.. This work is supported
by the BMAI project at IIS, the University of Tokyo.
\end{acknowledgements}

\bibliographystyle{spmpsci}      % mathematics and physical sciences
\bibliography{myref}

%-\begin{thebibliography}{}
%-\bibitem{RefJ}
%-Author, Article title, Journal, Volume, page numbers (year)
%-\bibitem{RefB}
%-Author, Book title, page numbers. Publisher, place (year)
% etc
%-\end{thebibliography}

\end{document}